\documentclass[preprint,preprintnumbers,amsmath,amssymb,amsfonts]{revtex4}
\usepackage{graphicx,bm,amssymb}
\usepackage[russian]{babel}

\begin{document}
\markboth{A.V. Shorokhov, K.N. Alekseev} {Theoretical backgrounds
of nonlinear THz spectroscopy of semiconductor superlattices}
\title{THEORETICAL BACKGROUNDS OF NONLINEAR THZ SPECTROSCOPY OF SEMICONDUCTOR SUPERLATTICES}
\author{ALEXEY V. SHOROKHOV}
\address{Institute of Physics and Chemistry, Mordovian State University, Saransk 430005, Russia\\
shorokhovav@math.mrsu.ru}
\author{KIRILL N. ALEKSEEV}
\address{Department of Physics, Loughborough University, \\
Loughborough LE11 3TU, United Kingdom\\
superlattice2@gmail.com}

 \maketitle
\section*{}
We consider terahertz absorption and gain in a single miniband of
semiconductor superlattice subject to a bichromatic electric field
in the most general case of commensurate frequencies of the probe
and pump fields. Using an exact solution of Boltzmann transport
equation, we show that in the small-signal limit the formulas for
absorption always contain two distinct terms related to the
parametric and incoherent interactions of miniband electrons with
the alternating pump field. It provides a theoretical background
for a control of THz gain without switching to the negative
differential conductivity state. For pedagogical reasons we
present derivations of formulas in detail.

\section{Introduction}
 We consider terahertz absorption and gain in a
single miniband of semiconductor superlattice subject to a
bichromatic electric field in the most general case of
commensurate frequencies of the probe and pump fields. Using an
exact solution of Boltzmann transport equation, we show that in
the small-signal limit the formulas for absorption always contain
two distinct terms related to the parametric and incoherent
interactions of miniband electrons with the alternating pump
field. It provides a theoretical background for a control of THz
gain without switching to the negative differential conductivity
state. For pedagogical reasons we present derivations of formulas
in detail.

There exists a great interest to generation, amplification and detection of coherent terahertz (THz) radiation
in semiconductor microstructures.\cite{Drag} It is stimulated by a rapid progress of THz sciences and technologies
ranging from the astronomy to medicine.\cite{Fer02,Ton07} Currently, the main direction in this active area of research
is the development of quantum cascade lasers\cite{Koh02} and amplifiers.\cite{QC-ampl} However, these devices require
relatively low temperature of operation.
\par
In theory, an intraminiband absorption of THz fields in dc biased semiconductor superlattices\cite{Esa70,Bass-rev,Wac02}
(SSLs) can be negative even at room temperature. It was predicted that this high-frequency gain arises in conditions of
negative differential conductivity (NDC).\cite{Kti72} However, a realization of THz amplifier or oscillator, based
on the NDC effect in SSL, is complicated by a development of space-charge instability\cite{Ridley63,Ign87} resulting in
a formation of high-field electric domains inside SSL. This NDC-induced electric instability is destructive for
THz gain.  To suppress the electric instability it was suggested a use of new types of SSLs, in particular lateral
superlattices,\cite{Fei05} a stack of short superlattices\cite{Sav04} or  miniband engineering.\cite{Rom04,Shmel08}
\par
Alternatively, a high-frequency gain can be achieved in SSLs
driven by strong ac fields (Refs.~\cite{Pav77}--\cite{Hya08}).
Negative absorption of a probe field arises if its frequency and
the frequency of ac pump field are commensurate.\cite{Pav77,Rom80}
In recent Letters\cite{Hya06,Hya07}, we clarified parametric
nature of this effect and found that it can exist without
switching to NDC in the time-average voltage-current
characteristic of SSL. This theoretical result allows to expect
that the undesirable electric instability can be effectively
suppressed in the case of parametric gain.
\par
Here we want to provide some mathematical details which have been omitted in the letter\cite{Hya07}.
Thus the aim of the present contribution is to derive the formulas describing absorption and gain of a THz probe field
in biased SSL in the presence of a strong THz pump field. Within the semiclassical approach we consider the most
general case of arbitrary commensurate pump $\omega_1$ and probe $\omega_2$ frequencies and take into account
effects of the relative phase $\varphi_0$ between the fields. We find that in the limit of a weak probe field the
absorption can be represented as a sum of the phase-dependent parametric and phase-independent incoherent terms.
In particular, our analysis provides a proof of the statement\cite{Hya07}
that a small-signal net gain of a probe field, which is not corrupted by an odinary generation of harmonics,
can exist in unbiased SSL only at even harmonics and in biased SSLs only at half-integer harmonics of the pump
electric field. Absorption at fractional harmonics is also considered.
Additionally, we compare our formula for small-signal absorption at even harmonics with
the corresponding formula derived by Pavlovich\cite{Pav77} in the particular case of unbiased SSL and $\varphi_0=0$.
We demonstrate that his formula formally contains an extra term, which nevertheless is identically zero.
Therefore, with account of this observation, we can confirm that the Pavlvich formula is correct.
For pedagogical reasons we present derivations of the main equations in this paper in detail.

\section{High-frequency absorption of an arbitrary probe}

We suppose that the total electric field $E(t)$ acting on
electrons  is a sum of the pump $E_{p}=E_0+E_1\cos(\omega_1t)$
($E_0$ is the dc bias) and probe $E_{pr}=E_2\cos(\omega_2t+\varphi_0)$ fields. In a real device, $E_{pr}$
may be a cavity mode tuned to the desired THz frequency.
We consider the most general case of commensurate frequencies
\begin{equation}
\label{freqs-def}
\frac{\omega_1}{\omega_2}=\frac{n}{m},
\end{equation}
where $n$ and $m$ are integers and $n/m$ is an
irreducible fraction. Dynamics of the
electrons belonging to a single miniband is well described by the
semiclassical approach\cite{Bass-rev,Wac02}  based on the use of Boltzmann
transport equation
\begin{equation}
\label{Boltzman}\frac{\partial f}{\partial t}+eE(t)\frac{\partial
f}{\partial p}=-\frac{f-f^{eq}}{\tau}
\end{equation}
together with the tight-binding dispersion relation
\begin{equation*}
\varepsilon(p)=\frac{\Delta}{2}\left[1-\cos\left(\frac{pd}{\hbar}\right)\right].
\end{equation*}
Here $\Delta$ is the miniband width, $d$ is the SSL
period, $p$ is the quasimomentum, $f^{eq}=\dfrac{d}{2\pi\hbar
I_0}\exp\left(\dfrac{\Delta}{2k_BT}\cos\dfrac{pd}{\hbar}\right)$ is
the equilibrium distribution function\cite{Ign87}, $I_0$ is the modified Bessel function of the argument
$\Delta/2k_BT$. The electron velocity, averaged over the distribution function $f(p,t)$, can be calculated as
\begin{equation}
\label{V-average-def}
\overline{V}(t)=\int V(p) f(p,t)\,  dp,
\end{equation}
where the integration is performed over Brillouin zone $|p|\leq \pi\hbar/d$, $V=\partial\varepsilon(p)/\partial p=V_0\sin(pd/\hbar)$ is the electron velocity, $V_0=\Delta d/2\hbar$ is the maximal electron velocity in the miniband.
We define the absorption $A(\omega_2)$ of the probe field $E_{pr}(t)$ as
\begin{equation}
\label{current}
A(\omega_2)=\langle\overline{V}(t)\cos(\omega_2t+\varphi_0)\rangle_t\equiv
\frac{1}{T}\int\limits_{0}^{T}\overline{V}(t)\cos(\omega_2t+\varphi_0)\,dt,
\end{equation}
where the time-averaging is performed over the common period $T=2\pi n/\omega_1=2\pi m/\omega_2$ of  both fields.
Note that the absorption depends on the relative phase $\varphi_0$.
Gain at frequecy $\omega_2$ corresponds to  $A(\omega_2)<0$.
Power $P$ absorbed ($P>0$) or emitted ($P<0$) in the miniband at frequency $\omega_2$ is proportional to $A(\omega_2)E_2$.
\par
The Boltzmann equation (\ref{Boltzman}) allows an exact solution (see
Appendix A). Using this solution we found the expression for absorption
in the form\footnote{The formula (\ref{gencur})  has been introduced in our preceding paper\cite{Sho06} for $\varphi_0$.}
\begin{eqnarray}
\label{gencur}
A(\omega_2)&=&\sum_{l_1,l_2=-\infty}^{\infty}\sum_{j=-\infty}^{\infty}
J_{l_1}(\beta_1)J_{l_2}(\beta_2)J_{l_1-jm}(\beta_1)\left[J_{l_2+jn-1}(\beta_2)+
J_{l_2+jn+1}(\beta_2)\right]\nonumber\\
&\times&\left[\frac{(\Omega_0+l_1\omega_1+l_2\omega_2)\tau\cos(jn\varphi_0)+\sin(jn\varphi_0)}
{1+(\Omega_0+l_1\omega_1+l_2\omega_2)^2\tau^2}\right],
\end{eqnarray}
where $J_n(x)$ is the odinary Bessel function, $\Omega_i=edE_i/\hbar$ ($i=0,1,2$)
and $\beta_i=\Omega_i/\omega_i$ ($i=1,2$) and $A$ is measured in the units
of peak velocity $V_p=V_0I_1(y)/2I_0(y)$ ($y=\Delta/2k_BT$)
corresponding to the peak current in Esaki-Tsu voltage-current characteristic\cite{Esa70,Bass-rev,Wac02}.
The derivation of Eq.~(\ref{gencur}) is presented in Appendix A.

\section{High-frequency absorption of a weak probe}

In what follows we will consider the small-signal limit $E_2\ll
E_1$. In this limit, we need to take only certain combinations of
indexes of Bessel functions in (\ref{gencur}). First, we consider
the most important case when the probe field is a harmonic of
the pump field.

\subsection{Absorption at harmonics ($\omega_2=m\omega_1$)}

In the limit of weak probe field ($\beta_2\ll1$), we need to take only the following combinations
of indexes of Bessel functions in (\ref{gencur}):
($l_2=0,j=\pm1$), ($l_2=0,j=\pm2$), ($l_2=\pm1,j=\mp2$),
($l_2=\pm1,j=0$). As a result we represent the absorption as a sum
of three terms
\begin{equation}
\label{A-general}
A=J^{\rm harm}(\varphi_0)+\frac{\beta_2}{2}(A^{\rm coh}+A^{\rm incoh})+O(\beta_2^2).
\end{equation}
Here $J^{\rm harm}$ depends on the pump amplitude ($\beta_1$) and relative phase $\varphi_0$ as
\begin{eqnarray}
\label{Aharm} J^{\rm
harm}(\varphi_0)&=&\sum_{l=-\infty}^{\infty}J_{l}(\beta_1)\left[J_{l-m}(\beta_1)\frac{(\Omega_0+l\omega_1)\tau\cos\varphi_0+\sin\varphi_0}
{1+(\Omega_0+l\omega_1)^2\tau^2}\nonumber\right.\\
&+&\left.J_{l+m}(\beta_1)\frac{(\Omega_0+l\omega_1)\tau\cos\varphi_0-\sin\varphi_0}
{1+(\Omega_0+l\omega_1)^2\tau^2}\right] .
\end{eqnarray}
Since $J^{harm}$ does not depend on a small amplitude of the probe
field this term is leading in Eq.~(\ref{A-general}). Note that
$J^{harm}$ is just the part of current at $m$-th harmonic which in
phase with the probe field. Importantly, negative sign of
$J^{harm}$ does not automatically mean a negative absorption of
the probe field and $P<0$, in contrast to the statement in
Ref.~\cite{litvinov}. Net gain can arise only if other terms in
the expression (\ref{A-general}) are negative.
\par
Next, $A^{\rm coh}$ describes the
parametric amplification of the probe field\cite{Hya07} due to its
coherent interaction with the pump field
\begin{eqnarray}
A^{coh}&=&\sum_{l=-\infty}^{\infty}J_{l}(\beta_1)\left[J_{l-2m}(\beta_1)
\frac{(\Omega_0+l\omega_1)\tau\cos(2\varphi_0)+\sin(2\varphi_0)}
{1+(\Omega_0+l\omega_1)^2\tau^2}
\right.\nonumber\\
&-&\left.J_{l+2m}(\beta_1)\frac{(\Omega_0+l\omega_1)\tau\cos(2\varphi_0)-\sin(2\varphi_0)}
{1+(\Omega_0+l\omega_1)^2\tau^2}\right]\nonumber\\
&-&2\sum_{l=-\infty}^{\infty}J_{l}(\beta_1)J_{l+2m}
(\beta_1)\frac{\sin(2\varphi_0)}{1+(\Omega_0+l\omega_1+\omega_2)^2\tau^2}
\end{eqnarray}
and $A^{\rm incoh}$ describes nonparametric effects
\begin{eqnarray}
\label{Aincoh}
A^{\rm incoh}=\sum_{l=-\infty}^{\infty}J_{l}^2(\beta_1)
\left[\frac{(\Omega_0+l\omega_1+\omega_2)\tau}
{1+(\Omega_0+l\omega_1+\omega_2)^2\tau^2}
-\frac{(\Omega_0+l\omega_1-\omega_2)\tau}
{1+(\Omega_0+l\omega_1-\omega_2)^2\tau^2}\right].\nonumber\\
\end{eqnarray}
It is important to notice that the sign of $A^{\rm incoh}$
plays an essential role in stabilization of space-charge
instability in SSL.\cite{Hya07,{Ale06}}
Namely, the necessary condition of electric stability is an operation at a positive slope of time-average
current-voltage characteristic. As a rule, this slope is positive if $A^{\rm incoh}>0$. Therefore it is desirable
to work in conditions which provide $A^{\rm incoh}>0$.

\subsubsection{Unbiased superlattice}

In the case of unbiased SSL ($E_0=0$) the harmonic
component of the current becomes
$$
J^{\rm harm}(\varphi_0)=
\left[1-(-1)^m\right]\sum_{l=-\infty}^{\infty}J_{l}(\beta_1)J_{l-m}(\beta_1)
\frac{l\omega_1\tau\cos\varphi_0+\sin\varphi_0}{1+l^2\omega_1^2\tau^2}.
$$
For odd $m$ the contribution of $J^{\rm harm}$ is larger than the sum of the coherent and incoherent components
of absorption. Therefore, the generation of odd harmonics can blur out a weaker effect of parametric gain.
In contrast, a generation of even harmonic is forbidden
due to symmetry, $J^{\rm harm}(\phi_0)=0$, and thus for even $m$
\begin{equation}
\label{A_net-nodc}
A=\frac{\beta_2}{2}(A^{\rm coh}+A^{\rm incoh})+O(\beta_2^2),
\end{equation}
where
\begin{eqnarray}
\label{Acoh0} A^{coh}&=&
2\sum_{l=-\infty}^{\infty}J_{l}(\beta_1)J_{l-2m}(\beta_1)
\frac{l\omega_1\tau\cos(2\varphi_0)+\sin(2\varphi_0)}{1+l^2\omega_1^2\tau^2}\nonumber\\
&-&2\sum_{l=-\infty}^{\infty}J_{l}(\beta_1)J_{l+2m}
(\beta_1)\frac{\sin(2\varphi_0)}
{1+(l\omega_1+\omega_2)^2\tau^2}
\end{eqnarray}
and
\begin{equation}
\label{Incoh0}
A^{incoh}=2\sum_{l=-\infty}^{\infty}J_{l}^2(\beta_1)
\frac{(l\omega_1+\omega_2)\tau}{1+(l\omega_1+\omega_2)^2\tau^2}.
\end{equation}
In our recent letter\cite{Hya07} we showed that the coherent absorption can be negative ($A^{coh}<0)$
at even harmonics due to a parametric resonance caused by Bragg reflections of miniband electrons.
Moreover, $|A^{\rm coh}|$ has a maximum at an optimal phase $\varphi_0=\varphi_{opt}$\cite{Hya07}.
In this case, the total absorption $A$ (\ref{A_net-nodc}) is determined by the concurrence of
$A^{\rm coh}<0$ and $A^{\rm incoh}>0$ and can be also negative (net gain) if  $|A^{\rm coh}|>A^{\rm incoh}$.
\par
We conclude this section noticing that Eqs. (\ref{Acoh0}) and (\ref{Incoh0}) have been first
obtained by Pavlovich.\cite{Pav77} The relations between our equations and formula
of Pavlovich are discussed in Appendix B.

\subsection{Absorption at integer half-harmonics ($\omega_2=m\omega_1/2$)}

In analogy with the preceding case, in the limit $\beta_2\ll1$ we
take only the certain combinations of indexes of Bessel
functions in (\ref{gencur}): ($l_2=0,j=\pm1$),
($l_2=\pm1,j=\mp1$), ($l_2=\pm1,j=0$). As a result we represent the absorption as a sum of two terms
\begin{equation} A(\omega_2)=\frac{\beta_2}{2}(A^{\rm coh}+A^{\rm incoh})+O(\beta_2^2).
\end{equation}
where
\begin{eqnarray}
\label{A-coh-half}
A^{coh}&=&\sum_{l=-\infty}^{\infty}J_{l}(\beta_1)\left[J_{l-m}(\beta_1)
\frac{(\Omega_0+l\omega_1)\tau\cos(2\varphi_0)+\sin(2\varphi_0)}
{1+(\Omega_0+l\omega_1)^2\tau^2}\right.\nonumber\\&-&
\left.J_{l+m}(\beta_1)\frac{(\Omega_0+l\omega_1)\tau\cos(2\varphi_0)-\sin(2\varphi_0)}
{1+(\Omega_0+l\omega_1)^2\tau^2}\right]\nonumber\\
&-&2\sum_{l=-\infty}^{\infty}J_{l}(\beta_1)J_{l+m}
(\beta_1)\left[\frac{1}
{1+(\Omega_0+l\omega_1+\omega_2)^2\tau^2}\right.\nonumber\\
&-&\left.\frac{1}
{1+(\Omega_0-l\omega_1-\omega_2)^2\tau^2}\right]\sin(2\varphi_0)
\end{eqnarray}
and
the incoherent absorption $A^{\rm incoh}$ is again given by Eq.~(\ref{Aincoh}).
\par
Analyzing (\ref{A-coh-half}) for $E_0=0$ we see that  $A^{coh}=0$ in the unbiased SSL and therefore
the total absorption becomes
\begin{equation}
\label{total_unbaised}
A(\omega_2)=\frac{\beta_2}{2}A^{\rm incoh},
\end{equation}
where $A^{\rm incoh}$ is given by (\ref{Incoh0}).
\par
Despite impossibility to have parametric gain at half-integer harmonics in unbiased SSL ($A^{\rm coh}=0$),
it is still feasible to have a negative absorption without electric instability -- {\it i.e.}
$A^{\rm coh}<0$, $A^{\rm incoh}>0$ and $|A^{\rm coh}|>A^{\rm incoh}$ -- in the case of dc biased
SSL.\cite{Hya07,Hyart-Cardiff}

\subsection{Absorption at fractional frequencies ($\omega_2=m\omega_1/n$ with $n\geq3$)}

First of all let us consider the case $n=3$. In the limit
$\beta_2\ll1$, we need to take only the following combinations of
indexes of Bessel functions in (\ref{gencur}): ($l_2=0,j=\pm1$),
($l_2=\pm2,j=\mp1$), ($l_2=\pm1,j=0$). As a result we can write
the total absorption as a sum of two terms
\begin{equation*}
A(\omega_2)=\frac{\beta_2}{2}A^{\rm incoh}+
\frac{\beta_2^2}{8}A^{\rm coh}+O(\beta_2^3),
\end{equation*}
where the incoherent absorption $A^{\rm incoh}$ is  given by Eq.~(\ref{Aincoh}) and
\begin{eqnarray}
\label{coh-frac-3}
A^{\rm coh}&=&\sum_{l=-\infty}^{\infty}J_{l}(\beta_1)\left[J_{l-m}(\beta_1)
\frac{(\Omega_0+l\omega_1)\tau\cos(3\varphi_0)+\sin(3\varphi_0)}
{1+(\Omega_0+l\omega_1)^2\tau^2} \right.\nonumber\\ &+&
\left.J_{l+m}(\beta_1)\frac{(\Omega_0+l\omega_1)\tau\cos(3\varphi_0)-\sin(3\varphi_0)}
{1+(\Omega_0+l\omega_1)^2\tau^2}\right]\nonumber\\
&+&\sum_{l=-\infty}^{\infty}J_{l}(\beta_1)\left[J_{l-m}
(\beta_1)\frac{(\Omega_0+l\omega_1-2\omega_2)\tau\cos(3\varphi_0)-\sin(3\varphi_0)}
{1+(\Omega_0+l\omega_1-2\omega_2)^2\tau^2}\right.\nonumber\\
&+&\left.J_{l+m}(\beta_1)\frac{(\Omega_0+l\omega_1+2\omega_2)\tau\cos(3\varphi_0)+\sin(3\varphi_0)}
{1+(\Omega_0+l\omega_1+2\omega_2)^2\tau^2}\right].
\end{eqnarray}
In particular case $E_0=0$ the incoherent absorption is (\ref{Incoh0}) and coherent absorption (\ref{coh-frac-3}) becomes
\begin{eqnarray}
A^{coh}&=&\left[1-(-1)^m\right]\sum_{l=-\infty}^{\infty}J_{l}(\beta_1)J_{l-m}(\beta_1)\times\nonumber\\
& & \left[ \frac{l\omega_1\tau\cos(3\varphi_0)+\sin(3\varphi_0)}{1+l^2\omega_1^2\tau^2}+
\frac{(l\omega_1-2\omega_2)\tau\cos(3\varphi_0)-\sin(3\varphi_0)}{1+(l\omega_1-2\omega_2)^2\tau^2} \right].
\end{eqnarray}
Hence, if $m$ is the even number $A^{coh}=0$ and the total absorption $A$ is essentially
determined by the nonparametric term $A=\beta_2 A^{\rm incoh}/2$.
\par
Second, we consider the case of $n>3$. In the limit of weak probe
$\beta_2\ll1$ one can obtain for the absorption
\begin{equation}
\label{nn} A(\omega_2)=\frac{\beta_2}{2}A^{\rm incoh}+
O(\beta_2^{n-1}).
\end{equation}
Finally, if the frequencies of the probe  and
pump fields are incommensurate, the expression for the absorption
$A(\omega_2)$ consists of only incoherent component $A^{incoh}$.

\section{Conclusion}

In summary, within the semiclassical approach we considered ac current in a
superlattice driven by a bichromatic electric field. We have obtained the formula for absorption (\ref{gencur})
in the most general case of commensurate frequencies of the probe and pump fields using an exact solution
of Boltzmann equation. We showed that in the small-signal limit
the absorption can always be represented as a sum of two distinct
terms describing the phase-depended parametric $A^{coh}$ and
phase-independent incoherent $A^{incoh}$ interactions of miniband
electrons with the high-frequency pump field. The incoherent
component $A^{incoh}$ describes the free carrier absorption
modified by the pump. The incoherent component $A^{coh}$ has a
parametric nature. As follows from Eq.~(\ref{nn}) in the case
$n>2$, the total absorption is  dominated  by the incoherent
absorption component. Hence it is possible to get
the parametric small-signal gain only in the cases when the probe field is
harmonic or half-harmonic of the pump field.
\par
Finally, we would like to notice that effects of nonlinear absorption, including the parametric gain,
are not limited to semiconductor superlattices and THz electric fields. These effects should be observable
in different physical systems demonstrating dissipative transport in a single energy band.

\section*{Acknowledgements}

We are grateful to Timo Hyart, Lauri Kurki, Natalia Alexeeva, Erkki Thuneberg and Feo Kusmartsev for
collaboration and useful discussions.
We especially thank Timo Hyart for his check of  formulas and valuable advices, including his comment related to
the physical meaning of Eq.~(\ref{Aharm}).
Sufficient part of this work was done in the University of Oulu, Finland. A.V.S. thanks Devision of Theoretical
Physics at Oulu for hospitality in 2005. Our research was partially supported by the Russian Programme ``Development of
Scientific Potential of Universities'', Academy of Finland and  AQDJJ Programme of European Science Foundation.

\appendix{Solution of Boltzmann equation}

In this Appendix we first represent the exact solution of the Boltzmann
equation (\ref{Boltzman}) in the case of total electric field
$E(t)=E_0+E_1\cos(\omega_1t)+E_2\cos(\omega_2t+\varphi_0)$ , and then derive the formula (\ref{gencur}).
\par
The electron distribution functions $f$ and $f^{eq}$ due to a
periodicity in the quasimomentum permit a representation in the
form of Fourier series
\begin{equation}
\label{distr} f(p,t)=\sum\limits_{m=-\infty}^{\infty}f_m(t)e^{i
m\varphi},\qquad
f^{eq}(p)=\sum\limits_{m=-\infty}^{\infty}f_m^{eq}e^{im\varphi},
\end{equation}
where $\varphi=pd/\hbar$.  The Fourier coefficients of the $f^{eq}$ are\cite{Ign87}
$f_m^{eq}=dI_m(y)/2\pi\hbar I_0(y)$, $I_m(y)$
is the modified Bessel function of the argument $y=\Delta/2k_BT$.
Substituting (\ref{distr}) in (\ref{Boltzman}) we get
\begin{equation}
\label{Bolf_m} \frac{\partial f_m}{\partial
t}-\left\{\Big[\Omega_0+\Omega_1\cos(\omega_1t)+
\Omega_2\cos(\omega_2t+\varphi_0)\Big]im-\frac{1}{\tau}\right\}f_m=
\frac{f_m^{eq}}{\tau}.
\end{equation}
Solving the homogeneous differential equation corresponding to (\ref{Bolf_m}),
we find
\begin{equation}
\label{f_m}
f_m=C(t)e^{im\Omega_0t}e^{-t/\tau}\exp\{im[\beta_1\sin(\omega_1
t)+\beta_2\sin(\omega_2 t+\varphi_0)]\}.\\
\end{equation}
To find the coefficient $C(t)$ we substitute this expression for $f_m$ in (\ref{Bolf_m}) and obtain
\begin{equation*}
C(t)=f_m^{eq}\sum_{l_1,l_2=-\infty}^{\infty}
J_{l_1}(m\beta_1)J_{l_2}(m\beta_2)\frac{\exp[(-im\Omega_0+1/\tau-il_1\omega_1-il_2\omega_2)t]}
{1-i(m\Omega_0+l_1\omega_1+l_2\omega_2)\tau}\exp(-il_2\varphi_0). \nonumber\\
\end{equation*}
Here we used the well-known Bessel formula\cite{Abr71}
\begin{equation}
\label{Bessel}
\exp(\pm i\beta\sin\theta)=\sum_{l=-\infty}^{\infty}J_l(\beta)\exp(\pm
il\theta).
\end{equation}
Therefore the expression for $f_m$ (\ref{f_m}) becomes
\begin{eqnarray}
f_m&=&f_m^{eq}\exp\{[\beta_1\sin(\omega_1t)+\beta_2\sin(\omega_2
t+\varphi_0)]im\}\nonumber\\
&\times&\sum_{l_1,l_2=-\infty}^{\infty}
J_{l_1}(m\beta_1)J_{l_2}(m\beta_2)\frac{\exp([-il_1\omega_1-il_2\omega_2]t)}
{1-i(m\Omega_0+l_1\omega_1+l_2\omega_2)\tau}\exp(-il_2\varphi_0) .\nonumber
\end{eqnarray}
Using again Eq.~(\ref{Bessel}) we get
\begin{eqnarray}
f_m&=&f_m^{eq}\sum_{l_1,l_2=-\infty}^{\infty}\sum_{k_1,k_2=-\infty}^{\infty}
J_{l_1}(m\beta_1)J_{l_2}(m\beta_2)J_{k_1}(m\beta_1)J_{k_2}(m\beta_2)\nonumber\\
&\times&\frac{\exp\{[-i(l_1-k_1)\omega_1-
i(l_2-k_2)\omega_2]t\}}{1-i(m\Omega_0+l_1\omega_1+l_2\omega_2)\tau}\exp[-i(l_2-k_2)\varphi_0].\nonumber
\end{eqnarray}
Let us change indexes $l_1-k_1=-\nu_1$, $l_2-k_2=-\nu_2$. Then
\begin{eqnarray}
\label{disfun}
f_m&=&f_m^{eq}\sum_{l_1,l_2=-\infty}^{\infty}\sum_{\nu_1,\nu_2=-\infty}^{\infty}
J_{l_1}(m\beta_1)J_{l_2}(m\beta_2)J_{l_1+\nu_1}(m\beta_1)J_{l_2+\nu_2}(m\beta_2)\nonumber\\
&\times&\frac{\exp[i(\nu_1\omega_1+
\nu_2\omega_2)t]}{1-i(m\Omega_0+l_1\omega_1+l_2\omega_2)\tau}\exp(i\nu_2\varphi_0).
\end{eqnarray}
Now we need to calculate the time-depended average electron velocity ({\it cf.} Eq.~(\ref{V-average-def}))
\begin{eqnarray}
\label{Vel}
\overline{V}(t)=\frac{\hbar}{d}\int\limits_{0}^{2\pi}V_0\sin(\varphi)f(\varphi)\,d\varphi.
\end{eqnarray}
Substituting the distribution function (\ref{disfun}) into
(\ref{Vel}), we obtain
\begin{eqnarray}
\label{Veloc}
\overline{V}(t)=\frac{\hbar}{d}V_0\sum_{m=-\infty}^{\infty}f_m\int\limits_{0}^{2\pi}\sin(\varphi)
\exp(im\varphi)\,d\varphi=\pi i\frac{\hbar V_0}{d}(f_1-f_{-1}).
\end{eqnarray}
Using the definition of the absorption (\ref{current}), (\ref{Veloc}), (\ref{disfun}) and denoting
\begin{eqnarray*}
G(m,l_1,l_2)=\frac{1}{1-i(m\Omega_0+l_1\omega_1+l_2\omega_2)\tau}=
\frac{1+i(m\Omega_0+l_1\omega_1+l_2\omega_2)\tau}
{1+(m\Omega_0+l_1\omega_1+l_2\omega_2)^2\tau^2},\nonumber\\
\end{eqnarray*}
we get
\begin{eqnarray}
\label{Ai}
&& A=\frac{i\pi\hbar
V_0}{2d}f_1^{eq}\sum_{l_1,l_2=-\infty}^{\infty}\sum_{\nu_1,\nu_2=-\infty}^{\infty}
J_{l_1}(\beta_1)J_{l_2}(\beta_2)J_{l_1+\nu_1}(\beta_1)J_{l_2+\nu_2}(\beta_2)\exp(i\nu_2\varphi_0)\nonumber\\
&\times&\left\{\left[G(1,l_1,l_2)-(-1)^{\nu_1+\nu_2}G(-1,l_1,l_2)\right]\langle
\exp[i(\nu_1\omega_1+\nu_2\omega_2+\omega_2)t]\rangle_t\exp(i\varphi_0)\right.\nonumber\\
&+&\left.\left[G(1,l_1,l_2)-(-1)^{\nu_1+\nu_2}G(-1,l_1,l_2)\right]\langle
\exp[i(\nu_1\omega_1+\nu_2\omega_2-\omega_2)t]\rangle_t\exp(-i\varphi_0)\right\}+\mathrm{c.c}\nonumber\\
&=&\frac{in_0V_0I_1
}{2I_0}\sum_{l_1,l_2=-\infty}^{\infty}\sum_{\nu_1,\nu_2=-\infty}^{\infty}
J_{l_1}(\beta_1)J_{l_2}(\beta_2)J_{l_1+\nu_1}(\beta_1)J_{l_2+\nu_2}(\beta_2)\nonumber\\
&\times&\left\{G(1,l_1,l_2)\langle
\exp[i(\nu_1\omega_1+\nu_2\omega_2+\omega_2)t]\rangle_t\exp[i\varphi_0(\nu_2+1)]\right.\nonumber\\
&-&\left.G^*(1,l_1,l_2)\langle
\exp[-i(\nu_1\omega_1+\nu_2\omega_2-\omega_2)t]\rangle_t\exp[-i\varphi_0(\nu_2-1)]\right\}+\mathrm{c.c.}
\end{eqnarray}
Performing the averaging in Eq.~(\ref{Ai}) over the common period $T=2\pi n/\omega_1=2\pi m/\omega_2$
of the pump and probe ac fields satisfying (\ref{freqs-def}), we obtain the conditions
\begin{equation}
\label{period}
 \nu_1\omega_1+\nu_2\omega_2+\omega_2=0,\qquad
\nu_1\omega_1+\nu_2\omega_2-\omega_2=0
\end{equation}
for the first and second term in (\ref{Ai}), respectively.
It follows from (\ref{period}) that
$\nu_1=-jm$ ($j\in\mathbb{Z})$, as well as that $\nu_2=jn-1$ for the first term  and $\nu_2=jn+1$ for
the second term in (\ref{Ai}), respectively. As a result we have
\begin{eqnarray}
A&=&\frac{iV_0I_1(y)
}{2I_0(y)}\sum_{l_1,l_2=\infty}^{\infty}\sum_{j=-\infty}^{\infty}
J_{l_1}(\beta_1)J_{l_2}(\beta_2)J_{l_1-jm}(\beta_1)\nonumber\\
&\times&\left[J_{l_2+jn-1}(\beta_2)
G(1,l_1,l_2)\exp(i\varphi_0jn)\right.\nonumber\\
&-&\left.J_{l_2+jn+1}(\beta_2)
G^*(1,l_1,l_2)\exp(-i\varphi_0jn)\right]+\mathrm{c.c.}
\end{eqnarray}
Taking into account that $V_p=V_0I_1(y)/2I_0(y)$ with $y=\Delta/2k_BT$,
\begin{equation*}
iG(1,l_1,l_2)\exp(i\varphi_0jn)=\frac{i-(\Omega_0+l_1\omega_1+l_2\omega_2)\tau}
{1+(\Omega_0+l_1\omega_1+l_2\omega_2)^2\tau^2}[\cos(jn\varphi_0)+i\sin(jn\varphi_0)]
\end{equation*}
and
\begin{equation*}
iG^*(1,l_1,l_2)\exp(-i\varphi_0jn)=\frac{i+(\Omega_0+l_1\omega_1+l_2\omega_2)\tau}
{1+(\Omega_0+l_1\omega_1+l_2\omega_2)^2\tau^2}[\cos(jn\varphi_0)-i\sin(jn\varphi_0)],
\end{equation*}
we obtain the final formula for absorption
\begin{eqnarray}
\label{absorp}
A&=&V_p \sum_{l_1,l_2=-\infty}^{\infty}\sum_{j=-\infty}^{\infty}
J_{l_1}(\beta_1)J_{l_2}(\beta_2)J_{l_1-jm}(\beta_1)\left[J_{l_2+jn-1}(\beta_2)+
J_{l_2+jn+1}(\beta_2)\right]\nonumber\\
&\times&\left[\frac{(\Omega_0+l_1\omega_1+l_2\omega_2)\tau\cos(jn\varphi_0)+\sin(jn\varphi_0)}
{1+(\Omega_0+l_1\omega_1+l_2\omega_2)^2\tau^2}\right] .
\end{eqnarray}

\appendix{Pavlovich formula}

In our notations the Pavlovich formula for absorption at harmonics
and integer-half harmonics in unbiased SSL (Eq.~(5) in
Ref.~\cite{Pav77}) can be written as
\begin{eqnarray}
\label{B1}
A&=&\beta_2\sum_{l=-\infty}^{\infty}J_{l}^2(\beta_1) \frac{(\nu-l)\omega_1\tau}{1+(\nu-l)^2(\omega_1\tau)^2}+
\frac{\beta_2}{2}\left[1+(-1)^{2\nu}\right] \nonumber\\
& \times & \sum_{l=-\infty}^{\infty}J_{l}(\beta_1)J_{l-2\nu}(\beta_1)\left[
\frac{l\omega_1\tau}{1+(l\omega_1\tau)^2}+
\frac{(\nu-l)(\omega_1\tau)}{1+(\nu-l)^2(\omega_1\tau)^2}\right],
\end{eqnarray}
where $\nu=\omega_2/\omega_1$.
It is easy to see that the first term in Eq.~(\ref{B1}) is the incoherent absorption for both $\nu=m/2$ and $\nu=m$ ({\it cf.} Eq.~(\ref{Incoh0})).
In the case of absorption at half-harmonics ($\nu=m/2$) that is the only nonzero term in the expression (\ref{B1}), what is consistent with our result
Eq.~(\ref{total_unbaised}). For absorption at harmonics ($\nu=m$) the second term in Eq.~(\ref{B1}) is just the coherent absorption
({\it cf.} Eq.~(\ref{Acoh0}) for $\phi_0=0$). Finally we consider the last term in (\ref{B1}) for $\nu=m$.
Introducing new summation index $k=2m-l$ we get
\begin{equation*}
\sum_{l=-\infty}^{\infty}J_{l}(\beta_1)J_{l-2m}(\beta_1)\frac{(m-l)(\omega_1\tau)}{1+(m-l)^2(\omega_1\tau)^2}=
-\sum_{k=-\infty}^{\infty}J_{k}(\beta_1)J_{k-2m}(\beta_1)\frac{(m-k)(\omega_1\tau)}{1+(m-k)^2(\omega_1\tau)^2}.
\end{equation*}
Therefore this term  equals to zero, as also easy to check numerically. We conclude that for $\phi_0=0$ our equations are identical to the Pavlovich formula.

\end{document}